\documentclass{eas}

\usepackage{graphicx}

\begin{document}

\topmargin=1truecm
\evensidemargin 1.8truecm
\oddsidemargin 1.8truecm

\TitreGlobal{Mass Profiles and Shapes of Cosmological Structures}




\hbox{\vspace{-3mm}\\
LPTENS-05/31}

\title{LIGHT \,DARK \,MATTER}

\author{Michel Cass\'e}\address{CEA/DSM/DAPNIA/SAp, Orme des Merisiers, and Institut d'Astrophysique de Paris.}

\author{\,Pierre Fayet}\address{Laboratoire de Physique Th\'eorique de l'ENS, UMR 8549 CNRS, Paris.}

%

\runningtitle{Light Dark Matter }

\setcounter{page}{1}


\index{M.Cass\'e}

\index{P.Fayet}

\index{Author3, C.}



%

\begin{abstract} The SPI spectrometer aboard of the INTEGRAL satellite
has released a map of the $\,e^+e^-\,$ annihilation emission line
of unprecedented quality, showing that most of the photons arise from a
region coinciding with the stellar bulge of the Milky Way. The
impressive intensity (\,$\simeq 10^{-3}$ photon cm$^{-2}$ s$^{-1}$) and
morphology (round and wide) of the emission is begging an explanation.

\vspace{.8mm}
Different classes of astrophysical objects could inject
positrons in the interstellar medium of the bulge, but the only
acceptable ones should inject them at energies low enough to avoid
excessive brems\-strahlung emission in the soft gamma ray regime. Among
the $\sim$ MeV injectors, none seems generous enough to sustain the high
level of annihilation observed. Even the most profuse candidate, namely
the $\beta^+$ radioactivity of $^{56}$Co nuclei created and expelled in
the interstellar medium by explosive nucleosynthesis of type Ia
supernovae, falls short explaining the phenomenon due to the small
fraction of positrons leaking out from the ejecta ($\,\approx 3\,\%$),
together with the low SNIa rate in the bulge ($\,\approx\, 0.03$ per
century).

\vspace{.8mm}
It is therefore worth exploring alternative solutions, as for instance,
the idea that the source of the positrons is the annihilation of light
dark matter (LDM) particles of the kind recently proposed, 
totally independently, by Bo$\!$ehm and Fayet. 
Assuming that LDM is the culprit, crucial
constraints on the characteristics (mass and annihilation cross-section)
of the associated particle may be discussed, combining direct gamma
ray observations and models of the early Universe. In particular,
the mass of the LDM particles should be significantly less than $100$ MeV, 
so that the $e^+$ and $e^-$  resulting from their annihilations 
do not radiate exceedingly through bremsstrahlung in the interstellar gas of the
galactic bulge. 

\end{abstract}

\maketitle

%




\section{Introduction}

\vspace{-2mm}

The 511 keV line arising from electron-positron annihilation is to
gamma-ray astronomy what the 21 cm line is to radioastronomy. The rather
young 511 keV gamma astronomy culminates with the recent release of a
splendid celestial map of the electron-positron line emission
(Kn\"odlseder {\em et al.} 2005\cite{Knodlseder2005}, and references therein), 
on the basis of data
collected by the SPI spectrometer aboard the INTEGRAL satellite
(Vedrenne {\em et al.} 2003\cite{Vedrenne2003}), showing a prominent bulge in the central
region of the galaxy. The interpretation of this map together with the
associated  energy spectrum around 511 keV is a major challenge of gamma
ray astronomy, going beyond, perhaps, pure astrophysics with its host of
energetic objects (millisecond pulsars, low mass X ray binaries, supernovae, hypernovae, gamma ray bursts and
so on). Since no astrophysical object or class of objects seems to be able to account for this large production rate
(1.5 10$^{43}$ positrons per second emitted in the galactic
bulge), it is natural to explore more exotic solutions.

\vspace{-.8mm}

\section{INTEGRAL/SPI spectra and maps}

\vspace{-2mm}

 It is gratifying to have at our disposal the map of the \,2\,$\gamma$ (the
511 keV thin line) and \,3\,$\gamma$ emissions (the continuum flanking this
line at lower energies) arising from $\,e^+e^-$ annihilations
through positronium, together with the detailed spectrum around 511
keV~ (Kn\"odlseder {\em et al.} 2005\cite{Knodlseder2005}; Jean {\em et al.} 2005\cite{Jean2005}). 
The spectral characteristics of the observed annihilation emission around 511 keV, 
i.e. a thin (\,$\approx$ \,1.3 keV) and intense line
overlying a broad line (\,$\approx$ \,5 keV) can be explained by positron
annihilations in the warm gas of the galactic bulge, assuming that
$e^+e^-$ annihilation is taking place mostly (more than 93 \%) 
via positrinonium (Jean {\em et al.} 2005\cite{Jean2005}; Churazov {\em et al.} 2005\cite{Churazov}), with 
a 25\,\% annihilation rate of the positronium into two photons. 
Most of the photons in this energy range (with a flux of
$\,\simeq 10^{-3}$ ph cm$^{-2}$ s$^{-1}$) originate in a volume
corresponding to the galactic bulge at a copious rate of $\,\simeq 10^{43}$
s$^{-1}$. More precisely, the extension of the emitting region,
cocentered to the galaxy, is $\approx\,$  9 degrees FWHM, assuming a Gaussian
distribution. Both the high injection rate and the quasi-spherical
geometry are difficult, if not impossible, to impute to known
astrophysical objects or phenomena.

\vspace{-.8mm}

\section{Potential astrophysical sources}

\vspace {-2mm}

Since the observed 511 keV emission is apparently diffuse (no evidence
for pointlike sources down to $\,\approx 10^{-4}$ ph cm$^{-2}$ s$^{-1}$), there are in
principle two general solutions to the problem of feeding the whole bulge with
positrons that annihilate within this volume. One may suppose, either
one or a few intense sources of low energy positrons with a large
diffusion coefficient of these charged particles in the magnetized
medium; or many sources (or even a continuum of sources) associated to a
small diffusion coefficient. The second solution is generally favoured
but the situation will remain unsettled until we can estimate safely the
diffusion coefficient of low energy electrons and positrons ($\,\sim\,$ MeV),
in a situation for which the quasi-linear diffusion theory ceases to apply since positrons
are no longer in resonance with Alfven waves (Parizot {\em et al.} 2005\cite{Parizot2005};
Jean {\em et al.} 2005\cite{Jean2005}).

\vspace{1mm}

{\bf 1.} High energy phenomena (cosmic rays, annihilation of heavy dark
matter particles, ... ) are excluded since together with positrons coming
from positive pion ($\,\pi^+$) decays, gamma rays of relatively high energy ($\,\sim$ 100 MeV
or so), arising from neutral pion ($\,\pi^\circ$) decays, should show up copiously, which
is definitively not the case. Indeed on the EGRET maps, we see no
counterpart to the bulge-like emission seen at 511 keV. This is an
important constraint, as we shall see in Section \ref{sec:brems} since it eliminates,
for instance, heavy neutralinos \,-- the favourite Dark Matter candidate of
particle physicists --\, as a possible source of these annihilating positrons.

\vspace{.3mm}

{\bf 2.} Moreover, SNIa fall short sustaining the high injection rate
(Cass\'e� {\em et al.} 2004\cite{Casse2004}; Prantzos 2004\cite{Prantzos2004};
Schanne {\em et al.} 2005\cite{Schanne2005}).  A
SNIa explosion rate of about 0.5 par century is required to sustain the
observed diffuse bulge emission considering that $\,\simeq 97\,\%$ of the
positrons produced by radioactivity are trapped in the ejecta, where
they annihilate (Milne {\em et al.} 1999\cite{Milne1999}), whereas a recent estimate based
on near infrared calibration (Mannucci {\em et al.} 2005\cite{Mannucci2005}) as well as on
the stellar mass of the old stellar system constituted by the galactic
bulge, leads to $\,\simeq\,0.03$ SNIa per century (Schanne {\em et al.} 2005\cite{Schanne2005}).
Hypernovae and the related 
gamma ray bursts (Cass\'e {\em et al.} 2004\cite{Casse2004};
Schanne {\em et al.} 2004\cite{Schanne2004}; Parizot {\em et al.} 2005\cite{Parizot2005})
 could do better,
but since the number of massive stars in the galactic disk (which are
their progenitors) is about ten times more than in the bulge, hot spots
of 511 keV emission in the disk should show up\,\footnote{Except if positrons escape rapidly
from the thin disk, where the gas is concentrated, which is quite
unlikely although not impossible.}, but this is not the case.

Also, according to Jean {\em et al.} (2005\cite{Jean2005}), it is doubtful that low
energy positrons from a single source in the central region of the
galaxy, including the central black hole, could fill the whole bulge, at
least on the basis of their estimate of the positron diffusion
coefficient (admittedly very uncertain). 

\vspace{.3mm}

{\bf 3.} Low mass X-ray binaries have also been suggested (Prantzos 2004\cite{Prantzos2004}), 
although no quantitative estimate has been made up to now. The same
argument holds for microquasars.

\vspace{.1mm}

Therefore, to say the least, the source of positrons in the bulge is subject
to intense debate.

\vspace{-.3mm}

\section{Light dark matter as the source of the bulge positrons}

\vspace{-1.4mm}

It is thus timely to propose an alternative solution and develop the idea
that annihilation of Light Dark Matter particles could be the required positron source
(Bo$\!$ehm {\em et al.}~2004b\cite{Boehmetal2004b}; Fayet 2004\cite{Fayet2004};
Cass\'e �{\em et al.}~2004\cite{Casse2004}). 
Indeed independently of INTEGRAL observations and before the release
of its first results light dark matter has been worked out by
Bo$\!$ehm \& Fayet (2004\cite{Boehmfayet2004}), and this special kind of dark matter has
shown to be perfectly adapted to explain the essential gamma ray data
relative to the bulge of the galaxy: intense
$\,e^+e^-$ annihilation signal without counterpart above 100 MeV.
Since it is light, its annihilation
cannot give rise to neutral pions and their decay products, high energy
photons.

\vspace{1mm}

Indeed, the emission of 511 keV photons is a two step process. First low
energy positrons are produced by the annihilation of light dark matter
(LDM) particles, then they propagate in the surrounding medium, they are
slowed down principally by ionisation losses, essentially form positronium, practically at
rest, and annihilate with electrons in 2 or 3 photons. 

\vspace{1mm}

Once posed the possible existence of LDM particles, it is worth studying the consequences
of this hypothesis on different astrophysical phenomena as for instance
Big Bang Nucleosynthesis (Serpico \& Raffelt 2004\cite{Serpico2004}) 
and the galactic or extragalactic gamma ray background 
(Beacom {\em et al.} 2005\cite{Beacom2005}; Ahn {\em et al.} 2005\cite{Ahn2005}; Rasera {\em et al.} 2005\cite{Rasera2005}) in view of constraining, as much as possible, 
the properties of the LDM particles. In particular, next to its mass and annihilation
cross-section, we would like to learn more about its nature, self-conjugate (i.e. being its own antiparticle) or not, bosonic or fermionic. 
Different versions of LDM particles have been proposed. 
In any case the introduction of light dark matter particles -- whether spin-0 or spin-$\frac{1}{2}$ -- 
has a price and requires the existence of new processes responsible for their annihilations.
These could correspond, in particular, to a new interaction mediated by a new neutral {\it \,light\,} 
spin-1 gauge boson, 
denoted by $U$, associated with an extension of 
the Standard Model gauge group to $\,SU(3)\times SU(2)\times U(1)\times \hbox{extra-}\,U(1)\,$ 
(Fayet 1981\cite{Fayet1981}; Bo$\!$ehm \& Fayet 2004\cite{Boehmfayet2004}; 
Fayet 2004\cite{Fayet2004}).

\vspace{1mm}

Another important issue concerns the possibility of deriving the dark
matter distribution through that of the 511 keV one (Bo$\!$ehm {\em et al.}
2004b\cite{Boehmetal2004b}; Bo$\!$ehm \& Ascasibar 2004\cite{Boehmascasibar2004};
Ascasibar {\em et al.} 2005\cite{Ascasibar2005}),
under the assumption that positrons annihilate close to their region of origin (on
the spot approximation), which seems safe for MeV energy positrons
(Jean {\em et al.} 2005\cite{Jean2005}), but becomes more doubtful above.

\vspace{1mm}

A crucial ingredient in this discussion is the velocity dependence of the annihilation
cross-section, also related to the mechanism by which the LDM particle couples to normal
fermions. This could be through the exchange of heavy (e.g. mirror) fermions in the case of a spin-0 LDM particle, 
or through the exchange of a light spin-1 $\,U$ boson (Boehm \& Fayet 2004\cite{Boehmfayet2004}; Fayet 2004\cite{Fayet2004}). 
Decomposing this cross-section $\,\sigma_{\rm ann}\,v_{\rm rel}\,$ as a sum \hbox{$\,a + b\,v^2$}, 
the question is to determine the parameters $a$ and $b$, which ponder the $S$-wave and $P$-wave contributions, 
in the language of particle physicists\,\footnote{More precisely $a$ vanishes if there is no $S$-wave contribution 
to the annihilation cross-section. Then $\,\sigma_{\rm ann}\,v_{\rm rel}\,\approx \,b\,v^2$, from the $P$-wave  contribution.}.
This is a considerable
program, related to the Dark Matter distribution in the central part of
the galaxy, unfortunately very uncertain, and the clumping of Dark
Matter around dense objects, even more uncertain. As a result we think that it is premature 
to draw definitive conclusions yet, 
and that, from this point of view, both $S$-wave and  $P$-wave dark matter annihilations within the Milky Way (or possibly a combination of them)
remain possible at this stage 
(Rasera {\em et al.} 2005\cite{Rasera2005}) -- of course with, in general, different dark matter distributions. 

\vspace{1mm}

A minimal requirement to qualify the Light
Dark Matter hypothesis is that it fulfils, at least, the following list
of stringent constraints:

\vspace{.5mm}

   {\bf 1.\ } It should amount to about 23\,\% of the present mass density of the
      Universe, to deserve the title of (non-baryonic) dark matter.
      
   {\bf 2.\ }  It should not jeopardize big bang nucleosynthesis.
   
   {\bf 3.\,} Its annihilation in the galactic bulge should explain the high positron
      in\-jection rate in the bulge, as implied by its 511 keV photon emission
      (intensity and morphology).
      
   {\bf 4.\ } The fast electrons and positrons released by the annihilations of LDM particles
      in the bulge should note radiate excessively through synchrotron,
      brems\-strahlung and inverse Compton processes --  which severely limits their
      injection energy, and hence the mass of LDM particles. 
      
   {\bf 5.\,}  Annihilation of light Dark Matter particles in the central part of all
      spheroidal systems (bulges of spiral galaxies and elliptical
      galaxies) must not lead to an excessively large contribution to the extragalactic
      soft gamma ray background.

 {\bf 6.\,}  And, last but not least, the new interaction responsible for the annihilation of LDM particles 
 should be both {\it sufficiently strong}, to ensure for sufficient annihilations of LDM particles
 (otherwise one would be led to a much too large contribution to the energy density of the Universe);
and {\it nevertheless discrete enough}, so that its effects could remain unnoticed up to now.

\vspace{1.5mm}

We focus here mainly on points 1 and 3, central to our purpose. 
Point 2 is thoroughly discussed by (Serpico \& Raffelt 2004\cite{Serpico2004}). 
Point 5 is
worked out separately (Rasera {\em et al.} 2005\cite{Rasera2005}). Point 4, still under study, is discussed in Section 
\ref{sec:brems}; already, on the basis of a simple calculation, a mass higher than
$100$ MeV is excluded by considering only the bremsstrahlung emission of
fast electrons and positrons in the interstellar medium of the galactic
bulge, and comparing it to the bulge emission below $100$ MeV.
Point 6 involves the fact that the amplitudes for the exchange of a $U$ boson 
between ordinary and Dark Matter particles are {\it much larger than 
weak-interaction amplitudes}, but only {\it at lower energies} 
\,-- for which weak interactions are really very weak --\,
while becoming smaller at higher energies.
Other constraints such as those associated with the possible production of a real spin-1 $\,U$ boson 
(that could behave very much like a spin-0 pseudoscalar axion, even if this one may be rendered 
to some extent ``invisible'' by increasing the extra-$U(1)$ symmetry-breaking scale, cf.~(Fayet 1980\cite{Fayet1980}, 1981\cite{Fayet1981})\,), or the parity-violation effects it might induce, 
are discussed in (Fayet 2004\cite{Fayet2004}; Bouchiat \& Fayet 2005\cite{Bouchiat2005}).

\section{Mass and annihilation cross-sections of LDM particles}

One of the pressing problems in modern cosmology is the nature of Dark
Matter. Understanding it is one of the most important challenges
of the century. Among many possibilities, WIMPs, with masses and
interaction cross-sections characterised by the weak scale, are generally
considered as the leading candidates. However it has been demonstrated
that dark matter should not be necessarily heavy (Bo$\!$ehm {\em et al.}
2004b\cite{Boehmetal2004b}; Fayet 2004\cite{Fayet2004}), 
provided that its interaction with matter at low energy is significantly stronger than 
for weak interactions, so as to ensure sufficient annihilations of LDM particles:
otherwise one would be led to a much too large contribution to the energy density of the Universe.

\vspace{1mm}
If this is the case, annihilation of Dark Matter particles -- and the corresponding
positron emission as well -- is strongly enhanced by the existence of this new channel, as compared to
classical weak interactions. This strengthening of annihilation has
beneficial effects on both the relic DM density and the 511 keV emission
of the galactic bulge, with some additional requirement on the energy
dependence of the cross-section. Indeed to avoid overproduction of 511 keV
photons in the bulge, the present annihilation cross-section ($\,\sigma_{\rm ann}\,v_{\rm rel}\,$) 
should be typically about four orders of magnitude lower than at the period of freeze
out of such LDM particles, in the early Universe. This implies (as considered on the basis of
a different argumentation by Bo$\!$ehm {\em et al.} (2004a)) that the cross-section 
must be velocity dependent, the velocity of DM particles being
about $\,0.4\,c\,$ at freeze out time, and $\,<10^{-3}\,c\,$ in the bulge, at the present
epoch. This is indeed the case if the new interaction is mediated by a new neutral gauge
boson, noted $\,U$, rather light, and very weakly coupled (at least to ordinary matter).

\section{Bremsstrahlung of fast electrons and positrons in the galactic bulge}
\label{sec:brems}


An interesting constraint on the mass of the LDM particle can be set from the bremsstrahlung emission 
of electrons and positrons released in the bulge by the annihilation of LDM particles. 
Actually, given the large rate of positrons produced, smaller dark matter masses tend to be favored, 
to avoid an excessive production of such gamma rays (Fayet 2004\cite{Fayet2004}; Cass\'e {\em et al.} 2004\cite{Casse2004}).
The inner bremsstrahlung emission of gamma rays (accompanying the electrons and positrons produced during dark matter radiative annihilation processes), in particular, forbids the larger values of the LDM mass,
restricting it to less than $\,\approx 20$ MeV$/c^2$ (Beacom {\em et al.} 2005\cite{Beacom2005}), 
or possibly more, depending on the exact evaluation of the radiative annihilation cross-section.

\vspace{1mm}

Independently of this process, at the energies of interest to us, below 100 MeV, the energy losses in the course of the 
propagation of the emitted electrons and positrons are dominated by ionisation (which ultimately results in heating the medium). The power injected in the ISM by these non-thermal leptons is totally negligible compared to that of stars under the form of photons (i.e. about 4 10$^{43}$ erg s$^{-1}$). While bremsstrahlung losses is a small fraction (decreasing at low energy) of the energy spent in ionising and heating the ambient gas, this emission, in the soft gamma ray regime, is indeed quite significant, as far as we are concerned here.
Let's suppose, for the simplicity of the argument, that the mass of the LDM particle is $100$ MeV$/c^2$. 
The annihilation of two of them releases an $e^+$ and an $e^-$ of $\,100$ MeV each. 
The positron should loose practically all its kinetic energy (about \,100 MeV $\simeq 1.6\,\ 10^{-4}$ erg)
to form positronium, almost at rest, before annihilating. It then leaves this kinetic energy to the ambient medium, 
and so does the associated electron.
The whole population of emitted positrons (1.5 \,10$^{43}$ s$^{-1}$), and associated electrons, 
should then leave a 
very large energy deposit of about 5 10$^{39}$ erg s$^{-1}\,$ altogether in the bulge volume (or about 5 10$^{38}$ erg s$^{-1}\,$, in the case of a 10 MeV$/c^2$ particle).

\vspace{1mm}

The relative fraction of energy lost by bremsstrahlung in the medium as compared to ionisation is density-independent, 
since both depend linearly on the gas (hydrogen) density of the medium traversed by the fast electrons and positrons, but very energy-dependent. This ratio bremsstrahlung/ionisation amounts to about $\,20\,\%$ at \,100 MeV (cf. Fig.~3 in Fatuzzio \&  Melia (2003\cite{Fatuzzio2003})) \,-- but not much more than \hbox{1 \%} at \,10 MeV.  
Concentrating on the higher energies, the resulting bremsstrahlung emission is already much more than what could be
allowed from the observations:  
i.e., less than 10$^{36}$ erg s$^{-1}$ in the 30 - 100 MeV energy band according to EGRET data, since there no hint of a bulge in the longitude and latitude profile of the 30-100 MeV gamma emission 
(Hunter {\em et al.} 1997\cite{Hunter1997})\footnote{The same argument can be opposed to the recent suggestion of positrons that would be injected at high energy by millisecond pulsar winds (Wang {\em et al.} 2005)\cite{Wang2005}.}.
\vspace{1mm}

This kind of argument, worked out more precisely (in progress) and applied to the gamma ray continuum measured at lower energies by the COMPTEL and IBIS-INTEGRAL missions, in the direction of the central region of the Milky Way, 
limits the mass of the LDM particle and constrains it to be significantly below 100 MeV.
The resulting limit is in fact potentially stronger than what may be obtained from the inner bremsstrahlung, 
as the rate of energy loss by usual bremsstrahlung in the medium is $\,\simeq 3 \%$ at 20 MeV, 
to be compared, in the case of the inner bremstrahlung, with a radiative
correction effect proportional to $\,\frac{\alpha}{\pi}\simeq 2\ 10^{-3}\,$.

\vspace{1mm}
Furthermore, in the favored case of a $P$-wave (or at least largely $P$-wave dominated at freeze out) annihilation cross-section, which is
much smaller in the Milky Way than at the time of freeze out, the intensity of the observed positron annihilation line also favors lower values of the LDM mass. Indeed, for a given dark matter mass distribution, the number density of dark matter particles ($\,n_{\rm dm}=\rho_{\rm dm}/m_{\rm dm}\,$) \,-- i.e. ultimately the number of positrons produced --\, is inversely proportional to the LDM mass. 
Having too heavy LDM particles, and therefore fewer of them, would then also necessitate postulating a rather large clumping factor for dark matter particles, to provide for the required annihilation flux of positrons.

\section {Conclusion}


Type Ia supernovae do not seem to inject enough positrons in the bulge ISM to explain the SPI-INTEGRAL observations. But, maybe, the SNIa rate is totally abnormal there for unknown reasons\,? or the galactic bulge does not behaves like a normal spheroidal system populated with old stars, as believed up to now\,?
If not, since compact objects seem to be unable to inject {\it low energy\,} positrons at a sufficient rate, one may have to accept the fact that something apparently rather exotic is happening in the middle of the galaxy.

\vspace{.5mm}
In this context, the usual hypothesis of heavy neutralinos (implied by supersymmetric extensions of the Standard Model, 
with $R$-parity conservation) is discarded as the source of such positrons, 
as it would also overproduce gamma rays in the 100 MeV - few GeV range swept by EGRET. 
\vspace{.5mm}

While light Dark Matter annihilation is not the only candidate, it deserves special consideration 
since it could explain, at the same time, the Dark Matter relic density and the $\,e^+e^-$  annihilation radiation in the central region of the galaxy; and at the moment, no physical or astrophysical test appears to contradict it. 
Laboratory research is thus essential to discover it.

\vspace{-.5mm}

\section {Acknowledgements}
\vspace{-1.5mm}

We thank Y. Rasera, J. Paul, S. Schanne and our colleagues in Saclay for interesting discussions.
















\end{document}